\documentclass{ws-procs9x6}

\begin{document}
\title{Self-tuning of the cosmological constant
\\
and Einstein Gravity on the brane 
\footnote{\uppercase{T}alk presented 
by \uppercase{B. Kyae} 
at {\it \uppercase{SUSY} 2003: 
\uppercase{S}upersymmetry in the \uppercase{D}esert}\/,
held at the \uppercase{U}niversity of \uppercase{A}rizona,
\uppercase{T}ucson, \uppercase{AZ}, \uppercase{J}une 5-10, 2003.
\uppercase{T}o appear in the \uppercase{P}roceedings. 
}}

\author{JIHN E. Kim}

\address{School of Physics, Seoul National University \\
Seoul, 151-747 Korea
}
\author{BUMSEOK KYAE,~~~Qaisar Shafi} 

\address{Bartol Research Institute, University of Delaware,\\
Newark, DE 19716 USA
}


\maketitle

\abstracts{
We show that a self-tuning mechanism of the
cosmological constant could work in 5D non-compact space-time.  
The standard model matter fields live only on the
4D brane. The change of vacuum energy on the brane  
%
just gives rise to dynamical shifts of the
profiles of the background metric and a bulk scalar field in the extra
dimension, keeping 4D space-time flat. 
%
%
To obtain the Newtonian potential on the brane at long distances, 
we introduce an additional
brane-localized 4D Einstein-Hilbert term.   
%
%
%
}

\section{Introduction and Summary} 

The problem of the cosmological constant or the absolute scale of
the potential energy has been considered as the most
serious mass hierarchy problem in modern particle
physics. In recent years, however, there has been a
hope to understand the vanishing cosmological constant in extra
dimensional field theories~\cite{original,kkl1,kachru}.
%
%
In the Randall-Sundrum-II (RS-II) type five dimensional (5D) theories
with a 3-brane embedded in a {\it non-compact} 5D space,
the 5D gravity sector contains a 5D cosmological constant $\Lambda_b$ and
a brane vacuum energy $\Lambda_1$.
For a flat four dimensional (4D) subspace, a fine-tuning relation
between $\Lambda_b$ and $\Lambda_1$
is inevitable in the RS-II model. 
However, by introducing a massless scalar field in the bulk,
it was shown that a self-tuning of the cosmological constant
is possible in the RS-II type setup~\cite{kkl1,original}.
To guarantee a massless scalar field, the scalar can be regarded  
as the dual field of a 5D 3 form tensor field $A_{MNP}$.  
%

%
%
%
To obtain $-\frac{1}{r}$ gravity potential,   
one might make the extra space compact by employing two branes.
However, the introduction of two branes accompanies
a fine-tuning relation between two brane cosmological
constants~\cite{nilles}. 
%
%
In this paper,
we employ only one brane, and  
introduce an additional brane-localized
4D Einstein-Hilbert term.    
As shown in Refs.~\cite{dvali,newDGP},
if one restricts matter fields only on the brane,
the $-\frac{1}{r}$ potential could be obtained   
even with an infinite volume extra dimension
by introducing a brane-localized gravity term.
%
%
%
%
%
In this paper, we will discuss only the $\Lambda_b=0$ case.  
%
%
\section{Self-tuning of the cosmological constant}  
We consider the following action in 5D space-time $(x^\mu,y)$,
\begin{eqnarray}
S&=&\int d^4xdy\bigg[\sqrt{|g_5|}\bigg(\frac{M_5^3}{2}R_5
-\frac{1}{2}\partial_M\phi\partial^M\phi+L_m^b\bigg)
\nonumber \\
&&~~~~~+\delta(y)\sqrt{|\bar{g}_4|}\bigg(\frac{M_4^2}{2}\bar{R}_4
+L_m^1-\Lambda_1\bigg) \bigg]~,~~\label{action}
\end{eqnarray}
%
%
where $\bar{R}_4$ is the 4D Einstein-Hilbert term 
$\bar{g}^{\mu\nu}\bar{R}^\rho_{\mu\rho\nu}$ 
constructed with a metric different from
the bulk metric $g_{MN}$:  
$\bar{g}_{\mu\nu}(x,y)=(1+\frac{\bar{\alpha}}{M_4^2}\partial_y^2)
[g_{\mu\nu}(x,y)\times\Omega^{-2}(x,y)]$   
with a dimensionless coupling $\bar{\alpha}$.   
%
%
At the linearized level, the $z$ derivative terms coming from
$\delta(y)M_4^2\sqrt{|\bar{g}_4|}\bar{R}_4$ maintain
the 4D gauge symmetry required for the spin-2 field~\cite{newDGP}.
A 4D general coordinate transformation still can be defined as
$\partial_z^2\bar{g}'_{\mu\nu}=\frac{\partial x^\rho}{\partial x^{'\mu}}
\frac{\partial x^\sigma}{\partial x^{'\nu}}\partial_z^2\bar{g}_{\rho\sigma}$
with $\partial_z^2(\frac{\partial x^\rho}{\partial x^{'\mu}})|_{z=0}=0$.
The ordinary standard model (SM) matter fields are assumed
to live only on the brane.  
The presence of brane terms with $\delta(y)$
in Eq.~(\ref{action}) explicitly breaks the 5D general covariance
into the 4D one at $y=0$. 
The introduction of $\delta(y)M_4^2\sqrt{|\bar{g}_4|}\bar{R}_4$
does not spoil any given symmetry~\cite{dvali,newDGP}.
We suppose that $M_4$ is comparable to $M_5$.
%
%
%
%
We introduce a $Z_2$ symmetry in
the extra space, and assign even (odd) parities to $g_{\mu\nu}$,
$g_{55}$ ($g_{\mu 5}$, $\phi$) under $y\leftrightarrow -y$.

The 4D flat background solutions turn out to be~\cite{original}   
\begin{eqnarray}
&&ds^2=\beta^2(y)~\eta_{\mu\nu}dx^\mu dx^\nu+dy^2 ~, 
\label{az1}\\
%
&&\beta(y)=\bigg[4a|y|+c\bigg]^{1/4}~, ~~\Omega(y)=\beta(y) ~, 
\\
&&\phi(y)=\frac{\sqrt{12M_5^3}}{4}~{\rm ln}\bigg[
\frac{4a}{c}|y|+1\bigg]\times{\rm sgn}(y)\label{phi3} ~, 
\label{sol3}
\end{eqnarray}
where 
%
$a$ ($>0$) is a dimensionful arbitrary constant.  
${\rm sgn}(y)$ is defined as $+1$ ($-1$) for $y>0$ ($y<0$). 
%
%
%
%
%
%
The integration constant $c$ should be
determined by the boundary conditions: 
\begin{eqnarray} \label{self1}
%
c=-\bigg(\frac{6M_5^3a}{\Lambda_1}\bigg)~>0 ~.    
%
\end{eqnarray}
To avoid naked
singularities at some points in the bulk, we should take only {\it positive}
values of $c$. 
Hence, the brane vacuum energy is required to be negative. 
In the boundary condition Eq.~(\ref{self1}), we note that $\Lambda_1$ is
never completely fixed by the Lagrangian parameters. Any arbitrary negative
value of $\Lambda_1$ allows a 4D flat space-time solution, by
adjusting $c$ and $a$ such that the boundary conditions at the brane
are fulfilled. Even if the electroweak and QCD phase transitions
change the brane vacuum energy $\Lambda_1\rightarrow\Lambda_1'$, a flat
4D space-time could be maintained by proper shifts
$c\rightarrow c'$ and/or $a\rightarrow a'$. It means that
the profiles of the metric and the scalar field in the extra
dimension are changed. Since they are dynamical fields, it is
always possible. 
We note that our self-tuning mechanism is not associated
with any parameter sensitive to low energy physics. 
As will be shown, the 4D Newtonian constant
is given by $M_4$ rather than $M_5$ in our model.

According to Ref.~\cite{hawking},
the flat universe is quantum mechanically most probable in 4D space-time,
even though universe with a non-zero 4D curvature
is also classically allowed.
Hence, the flat universe can always be {\it dynamically} chosen
through the self-tuning mechanism.

\section{Metric perturbation and Conclusion}

The brane-localized gravity term does not affect
the self-tuning mechanism, but it is essential 
in gravity interaction on the brane.     
Let us consider metric perturbation near the background solutions.   
For convenience of the further analysis,
we change the $(x,y)$ coordinate into the conformal coordinate $(x,z)$.
With the gauge choices $\partial^\mu
(h_{\mu\nu}-\frac{1}{2}\eta_{\mu\nu}h)=h_{\mu 5}=0$,  
the $(\mu\nu)$ component of the linearized Einstein equation turns out 
to be~\cite{original} 
\begin{eqnarray}
&&-\frac{1}{2}\bigg[\bigg(\nabla_5^2-3\frac{K'}{K}\partial_z\bigg)
+\frac{M_4^2}{M_5^3}\delta(z)K\nabla_4^2\bigg(1+\frac{\alpha}{M_4^2}
\partial_z^2\bigg)\bigg]\bigg[h_{\mu\nu}
-\frac{1}{2}\eta_{\mu\nu}h\bigg] ~~~~
\nonumber \\
&&~~~~~=\frac{1}{6}\bigg(
\partial_\mu\partial_\nu-\eta_{\mu\nu}\nabla_4^2\bigg)\bigg[f-h^0\bigg]
+\frac{1}{M_5^3}T^b_{\mu\nu}+\frac{1}{M_5^3}\delta(z)KT^1_{\mu\nu} ~,
\label{simplemn}
\end{eqnarray}
where $K(z)=\frac{1}{\beta(y(z))}=(3a|z|+c^{3/4})^{-1/3}$ with  
$|z|=\int^z_0 dy/\beta(y)
=\frac{1}{3a}[(4a|y|+c)^{3/4}-c^{3/4}]$.   
$'$ denotes the derivative with respect to $z$, and
$\nabla_5^2$ and $\nabla_4^2$ indicate
$\eta^{\mu\nu}\partial_{\mu}\partial_{\nu} +\partial_z^2$ and
$\eta^{\mu\nu}\partial_{\mu}\partial_{\nu}$, respectively.      
$h$ is defined as $\eta^{\mu\nu}h_{\mu\nu}$, and $h^0(x)\equiv h(x,z=0)$.   
%
%
The SM matter fields only contribute to $T^1_{\mu\nu}$.
For simplicity, we assume that another massless bulk scalar field
independent of $z$ 
makes a dominant contribution to $T^b_{MN}$. 
%
%
$f(x)$ satisfies $\nabla_4^2f(x)=\frac{2}{M_5^3}T^b(x)$. 
Indeed, gravity interaction on the brane is insensitive to $T^b_{MN}$
in our model~\cite{newDGP}.
%
%
%
In Eq.~(\ref{simplemn}) we note that at $z=0$, 
the equation of motion for the trace mode is
$\nabla_4^2h^0(x)=\frac{2}{M_4^2}T^1(x)$, whereas at $z\neq 0$, 
%
it is $[\nabla_5^2-3\frac{K'}{K}\partial_z](h-h^0)=0$.
The bulk solution of Eq.~(\ref{simplemn}) in 4D momentum space $(p,z)$ 
is given by Bessel functions of order zero~\cite{original}.  

The terms with the coupling $\alpha$ in Eq.~(\ref{simplemn}) 
require to satisfy a non-trivial boundary condition at $z=0$: 
%
$\partial_z\tilde{h}_{\mu\nu}|_{z=0^+}
=\partial_z\tilde{h}_{\mu\nu}|_{z=0^-}=0$~\cite{newDGP}.
Otherwise highly singular terms proportional to $\delta^2(z)$ arise, 
which can not be matched to the right handed side of Eq.~(\ref{simplemn}).  
%
%
%
Then, since $\partial^2_zh_{\mu\nu}$, $\partial^2_zh$ do not
induce any delta function,  
we have only to compare the coefficients of the delta functions
appearing in Eq.~(\ref{simplemn}) to fulfill the boundary condition at $z=0$.  
%
%
%
%
We found the graviton solution at $z=0$ is given by~\cite{original}   
\begin{eqnarray} \label{final}
\tilde{h}_{\mu\nu}(p,z=0)=\frac{2}{M_4^2p^2}\bigg[
\tilde{T}^{1}_{\mu\nu}(p)-\frac{1}{2}\eta_{\mu\nu}\tilde{T}^1(p)\bigg]
+O\bigg(\frac{\alpha\tilde{T}^{1T}_{\mu\nu}}{M_4^4},
\frac{\alpha \tilde{S}^T_{\mu\nu}}{M_5^3M_4^2}\bigg) ~,  
\end{eqnarray}
where $\tilde{T}^1\equiv \eta^{\mu\nu}\tilde{T}^1_{\mu\nu}$, and   
$\tilde{S}_{\mu\nu}\equiv -\frac{M_5^3}{6}(p_\mu p_\nu
-\eta_{\mu\nu}p^2)[\tilde{f}-\tilde{h}^0]+\tilde{T}^b_{\mu\nu}$.   
The unspecified part of order $\alpha$ are negligible at low energies.  
Hence, at low energies 
$\tilde{h}_{\mu\nu}$ in Eq.~(\ref{final})
reproduces the same result that the 4D Einstein gravity theory predicts 
with $G_N\equiv 1/(8\pi M_4^2)$.    
%

In conclusion, as we have shown, 4D flat solution is always possible
independent of the magnitude of 4D brane cosmological constant
in 5D non-compact space-time, through dynamics 
by the 5D gravity and a massless scalar.   
Since the SM fields are localized on the brane,
4D vacuum energy by them and its variation do not destroy the 4D flatness.
An additional brane-localized gravity term 
successfully induces $-\frac{1}{r}$ potential on the brane.
%
%
%


\end{document}